\documentclass[aps,prb,10pt,twocolumn,superscriptaddress]{revtex4-1}

\usepackage{graphicx}
\usepackage{SIunits}
\usepackage{color}

\begin{document}

\title{Li--intercalated Graphene on SiC(0001): an STM study}

\author{Sara Fiori}
\affiliation{NEST, Istituto Nanoscienze--CNR and Scuola Normale Superiore, Piazza San Silvestro 12, 56127 Pisa, Italy}

\author{Yuya Murata}
\affiliation{NEST, Istituto Nanoscienze--CNR and Scuola Normale Superiore, Piazza San Silvestro 12, 56127 Pisa, Italy}

\author{Stefano Veronesi}
\affiliation{NEST, Istituto Nanoscienze--CNR and Scuola Normale Superiore, Piazza San Silvestro 12, 56127 Pisa, Italy}

\author{Antonio Rossi}
\affiliation{NEST, Istituto Nanoscienze--CNR and Scuola Normale Superiore, Piazza San Silvestro 12, 56127 Pisa, Italy}
\affiliation{Center for Nanotechnology Innovation @NEST, Istituto Italiano di Tecnologia, Piazza San Silvestro 12, 56127 Pisa, Italy}

\author{Camilla Coletti}
\affiliation{Center for Nanotechnology Innovation @NEST, Istituto Italiano di Tecnologia, Piazza San Silvestro 12, 56127 Pisa, Italy}

\author{Stefan Heun}
\email{stefan.heun@nano.cnr.it}
\affiliation{NEST, Istituto Nanoscienze--CNR and Scuola Normale Superiore, Piazza San Silvestro 12, 56127 Pisa, Italy}

\date{\today}

\begin{abstract}
We present a systematical study via scanning tunneling microscopy (STM) and low-energy electron diffraction (LEED) on the effect of the exposure of Lithium (Li) on graphene on silicon carbide (SiC). We have investigated Li deposition both on epitaxial monolayer graphene and on buffer layer surfaces on the Si--face of SiC. At room temperature, Li immediately intercalates at the interface between the SiC substrate and the buffer layer and transforms the buffer layer into a quasi--free--standing graphene. This conclusion is substantiated by LEED and STM evidence. We show that intercalation occurs through the SiC step sites or graphene defects. We obtain a good quantitative agreement between the number of Li atoms deposited and the number of available Si bonds at the surface of the SiC crystal. Through STM analysis, we are able to determine the interlayer distance induced by Li--intercalation at the interface between the SiC substrate and the buffer layer.
\end{abstract}

\maketitle

\section{Introduction}

Graphene is a one--atom thick carbon sheet arranged in a honeycomb structure. Its extraordinary properties, including a large specific surface area, excellent electrical and thermal conductivity, high charge mobility, great mechanical strength, low optical absorbance and density, and unusual flexibility make it one of the most studied materials at the moment.\cite{art:49} Its applications range over the most varied fields: hydrogen\cite{art:75,art:43,art:56,art:70,rossi2015nano,Takahashi2016} and energy storage,\cite{art:45} high--frequency electronics,\cite{Tassin2013} photodetectors,\cite{Mueller2010,Vicarelli2012} biology,\cite{chung2013biomedical,Kostarelos2014,Heerema2016} chemistry,\cite{Lv2014} and many other fields.\cite{Novoselov2012,Radha2016} 

In several cases, the already extraordinary properties of graphene can be enhanced by interaction with other species of the periodic table. \cite{art:12} Among these, Li--functionalized graphene has attracted a huge interest from a theoretical,\cite{art:11} experimental,\cite{art:13,art:14} and also technological\cite{art:45} point of view. The study of this compound is encouraged by its potential for batteries,\cite{art:45,shan2014visualizing} but also as doping material for graphene devices for superconductivity\cite{art:46,tiwari2015superconductivity,ludbrook2015evidence} and hydrogen storage.\cite{art:46,art:47} The experimental interest in Li--functionalized graphene is quite new, and there are lots of features to investigate. 

Epitaxial monolayer graphene (EMLG), due to the growth process, is formed by a graphene sheet stacked above a buffer layer (BL) (a carbon layer at the interface with the SiC substrate), which is partially bound to the SiC substrate, and these covalent bonds make it corrugated.\cite{emtsev2008interaction,art:1} This involves, for the BL, the loss of all graphene electronic properties and compromises the maximum efficiency of epitaxial graphene devices. In order to resolve this issue, the detachment of the BL from the SiC substrate and its conversion to quasi--free--standing monolayer graphene (QFMLG) \cite{riedl2009quasi} is desirable, and the experimental results already obtained from Li--doped graphene show such detachment.\cite{art:13,art:14,art:15} Li--intercalation below the graphene surface has already experimentally been observed by Low Energy Electron Microscopy (LEEM) and LEED.\cite{art:13,art:14,art:37} In such a way, the buffer layer is converted to QFMLG. This has stimulated the scientific interest to study in more detail the interaction between Li and epitaxial graphene on SiC.

This paper is the first work in which the interaction between Li atoms and graphene is studied in detail by STM. We show that Li readily intercalates at room temperature (RT), both on epitaxial monolayer graphene and on the buffer layer. Finally, we report what happens when annealing cycles are performed on these samples. All reported results have been obtained by STM and LEED in ultra high vacuum (UHV) environment.

\section{Methods}

Epitaxial monolayer graphene on silicon carbide (G/SiC) was obtained by annealing 6H-SiC(0001) samples in a high-temperature BM reactor (Aixtron) under an Ar atmosphere at about 1400~$^\circ$C and $780$~mbar. Samples with a mixture of monolayer graphene and BL surfaces are obtained, allowing a complete overview of the interaction between Li and graphene. The quality, homogeneity, and precise thickness of the graphene were assessed by Raman spectroscopy. LEED and STM characterization were performed in UHV environment. 

\begin{figure*}[t]
\includegraphics[width=\textwidth]{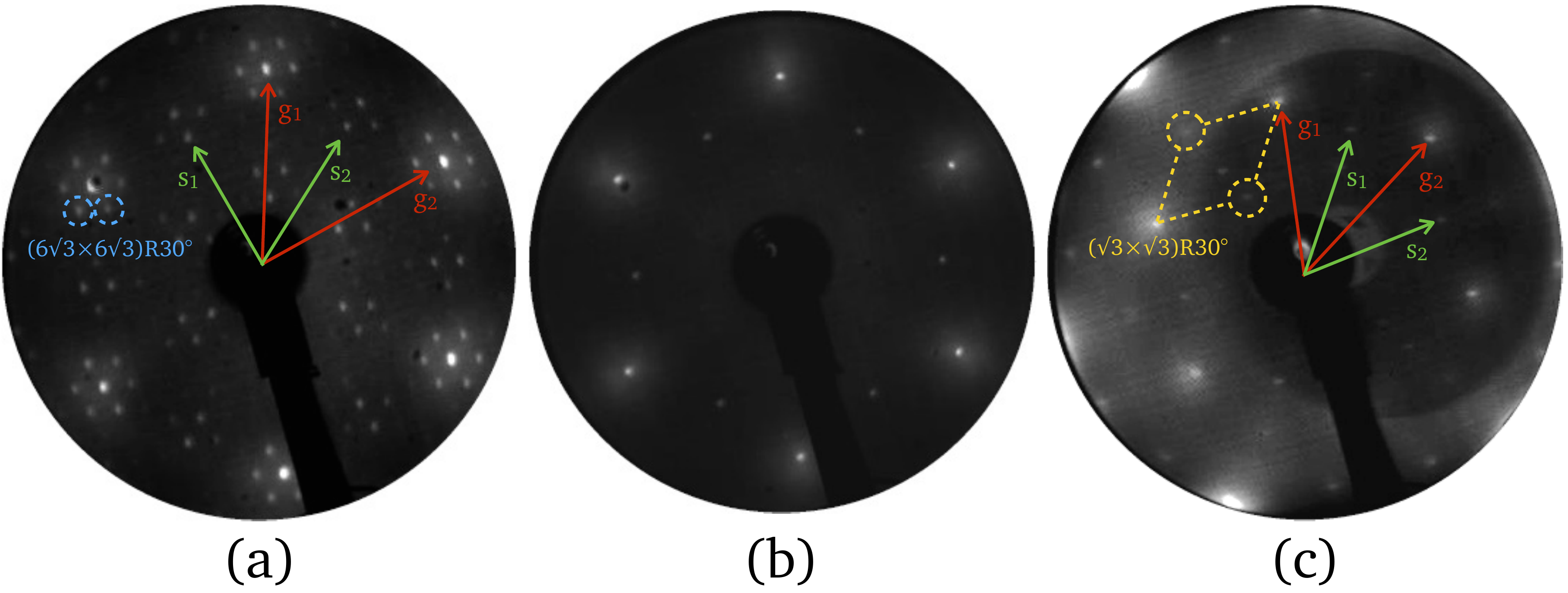} 
\caption{\label{fig: fig1}LEED diffraction patterns for (a) pristine graphene, and after having deposited (b) $0.28 \pm 0.02$~ML of Li, and (c) $0.56 \pm 0.04$~ML of Li. Electron energy: (a) 92.5 eV, (b) 95.5 eV, (c) 135.6 eV.}
\end{figure*}

In order to perform the Li deposition on EMLG, a Li evaporator was homebuilt using commercial Li dispensers from SAES Getters. During Li evaporation in the preparation chamber of the STM UHV system, the pressure was always kept below $6 \times 10^{-10}$~mbar. All Li depositions on G/SiC have been performed at RT. To calibrate the Li dispenser, a Si(111)--$\left( 7 \times 7 \right)$ reconstructed sample was used. Li was deposited on the Si(111)--$\left( 7 \times 7 \right)$ surface which was then annealed at 450~$^\circ$C, which resulted in a modification of the surface reconstruction to a Li--induced $\left( 3 \times 1 \right)$ surface.\cite{art:22,art:36} This has allowed to quantify the amount of Li deposited on the Si surface in a given deposition time, resulting in a calibration coefficient of $1.18 \times 10^{14}$~Li atoms/cm$^2$/min. Considering the C atom density in a graphene sheet (3.817$\times 10^{15}$~C atoms/cm$^2$), we define a Li coverage of 1 monolayer (ML) on G/SiC as formed by one Li adatom for each C atom of the graphene film. Based on this definition, we obtain a Li deposition rate of $0.031  \pm 0.002$~ML/min on G/SiC. 

The scanning tunneling microscope used to perform the experiment is a variable-temperature (VT) ultra high vacuum STM from RHK Technologies with a base pressure of $3 \times 10^{-11}$~mbar. The tungsten tips used to scan the samples were electrochemically etched, degassed in situ, and then flashed by applying high voltage. The samples were degassed in situ by direct current heating. The annealing temperature, monitored by an IR pyrometer and a thermocouple, was kept to 500 $^\circ$C overnight. All LEED and STM measurements were performed at RT. The STM images were taken in constant-current mode using several values for bias and current. The WSxM software package was used to analyze STM images.\cite{horcas2007wsxm}

\section{Results}

To follow the Li deposition process on G/SiC, we performed small Li deposition steps of $0.016 \pm 0.001$~ML at RT, observing, by LEED, progressive changes in the diffraction patterns. LEED diffraction patterns obtained for pristine graphene, and after deposition of $0.28 \pm 0.02$~ML and $0.56 \pm 0.04$~ML of Li are reported in Figs.~\ref{fig: fig1}(a), (b), and (c), respectively. Figure~\ref{fig: fig1}(a) shows the well--known LEED diffraction pattern of pristine graphene.\cite{art:1} It consists of three types of spots: (i) a six-fold pattern from graphene which has a honeycomb atomic configuration; (ii) the SiC--$\left( 1 \times 1 \right)$ spots from the bulk SiC substrate attenuated by the graphene film; (iii) a large number of spots from the $\left( 6 \sqrt{3} \times 6 \sqrt{3} \right)$R30$^\circ$ reconstruction. The red arrows labelled {\itshape g$_{1,2}$} in Fig.~\ref{fig: fig1}(a) and (c) indicate the graphene reciprocal lattice vectors, while the green arrows labelled {\itshape s$_{1,2}$} indicate those of the SiC. The other spots, indicated by blue circles in Fig.~\ref{fig: fig1}(a), correspond to the $\left( 6 \sqrt{3} \times 6 \sqrt{3} \right)$R30$^\circ$ reconstruction of the BL.\cite{emtsev2008interaction} Such moir\'{e} pattern arises from the difference in the lattice constants between SiC and graphene, which produces a mismatch between the graphene and SiC unit cells with rotation of 30$^\circ$.

In Fig.~\ref{fig: fig1}(b), the diffraction pattern shows only the $\left( 1 \times 1 \right)$ structures of graphene and SiC. The moir\'{e} spots around the graphene and SiC spots have completely disappeared, indicating a complete detachment of the BL from the substrate. This is clear evidence for an intercalation of Li to the interface between SiC and the BL.

Depositing more Li ($0.56 \pm 0.04$~ML), the LEED diffraction pattern shows a modification. The periodicity of the graphene surface changes from a $\left( 1 \times 1 \right)$ to a $\left( \sqrt3 \times \sqrt3 \right)$R30$^\circ$ superstructure. The new superstructure is related to the graphene lattice, as highlighted by the yellow dashed parallelogram in Fig.~\ref{fig: fig1}(c). The well-known spots related to graphene and SiC structures are still visible, indicated by red and green arrows, respectively, while the $\sqrt3$ spots are indicated by yellow circles. These experimental results are in good agreement with literature.\cite{art:13,art:14,art:104,kanetani2012intercalated} The $\left( \sqrt 3 \times \sqrt 3 \right)$R30$^\circ$ superstucture has been associated with Li--intercalation in between the two graphene layers.\cite{art:16} In excellent agreement with these LEED data, our STM experimental results presented in the following show how Li atoms interact with monolayer graphene and BL surfaces.

\subsection{Li on epitaxial monolayer graphene}

\begin{figure}[t]
\includegraphics[width=\columnwidth]{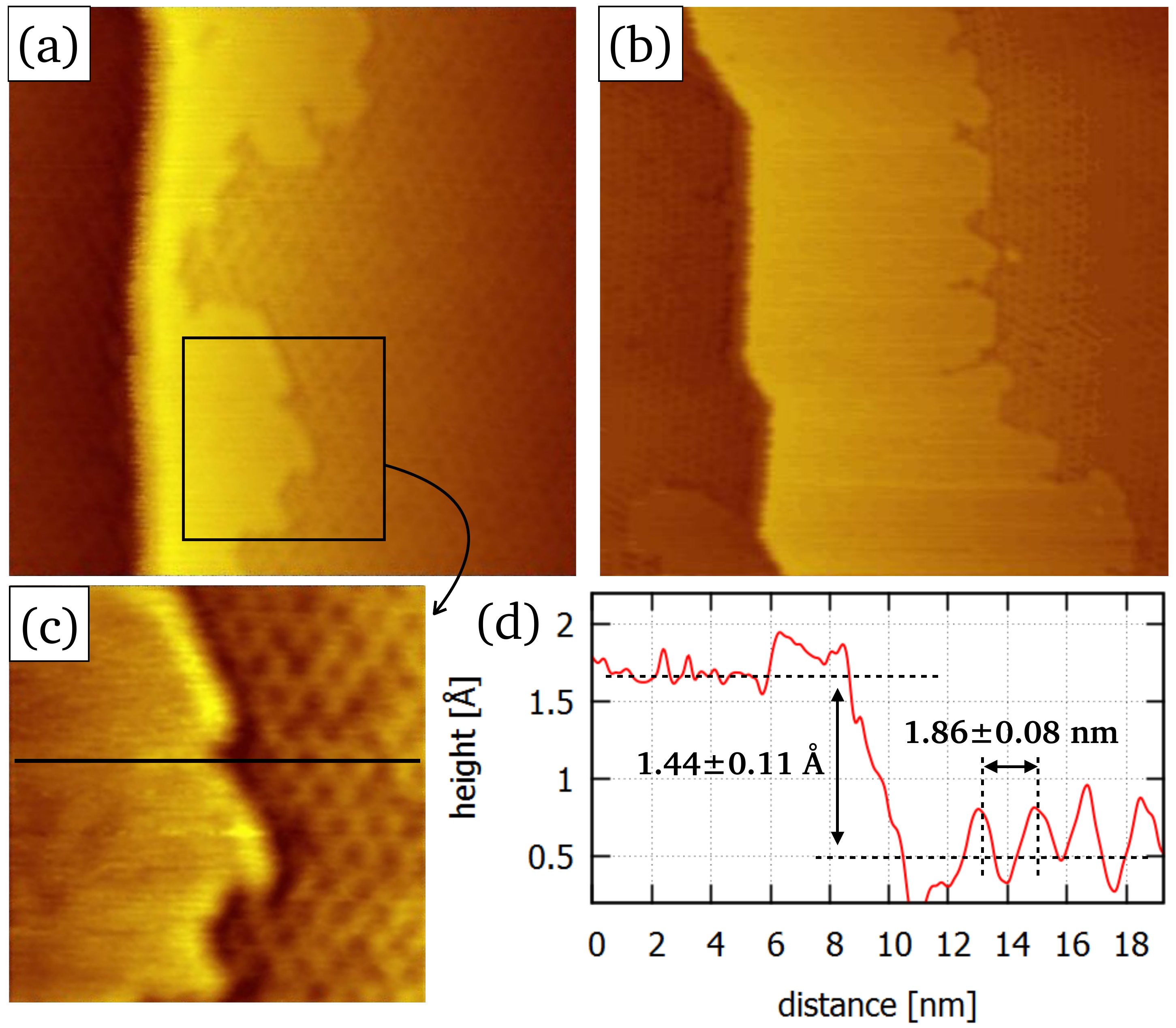}
\caption{\label{fig: fig2}(a) STM topographical image taken after deposition of 0.031~ML Li on EMLG. The moir\'{e} pattern is visible in the left and right parts of the image while in the middle there is a flat and not reconstructed region. Scan area: $50 \times 50$~nm$^2$. (b) STM topographical image taken after deposition of 0.047~ML Li on EMLG. The image shows that the Li--intercalated area has grown as compared to (a). Scan area: $72 \times 72$~nm$^2$. (c) Magnification taken from the area indicated by the solid square in (a). Scan area: $20 \times 20$~nm$^2$. (d) Cross sectional plot taken along the black line in (c). Image parameters: (a) $-$500 mV, $-$170 pA; (b) 1 V, 1~nA; (c) $-$500 mV, $-$170 pA.}
\end{figure}

After deposition of $0.031 \pm 0.002$~ML of Li on EMLG at RT, a clear morphological change of the surface is visible through the STM topographical image reported in Fig.~\ref{fig: fig2}(a). The image displays a $50 \times 50$~nm$^2$ scan area in which a step in the SiC substrate separates a lower (left, dark) from a higher (right, bright) terrace. On the right terrace, two different domains are visible: the first is along to the SiC step edge and extends inwards separated from the second flat and uniform region by a jagged boundary. The color contrast suggests that the region along the steps edge is slightly higher than the rest of the terrace. To understand the difference between these areas, we measured a zoom-in image, displayed in Fig.~\ref{fig: fig2}(c). We note that the right part shows a periodic structure (consistent with a $\left( 6 \sqrt{3} \times 6 \sqrt{3} \right)$R30$^\circ$ reconstruction) that is not present in the left part of the image, i.e.~along the step edge. A cross sectional plot, shown in Fig.~\ref{fig: fig2}(d), provides quantitative information: (i) we measure a step height between the two domains of $1.44 \pm 0.11$~\angstrom. The step height does not significantly vary over several measurements taken on different samples and under different experimental conditions such as different bias voltages. The average step height is $1.6 \pm 0.2$~\angstrom. This value is clearly lower than the SiC bilayer step height ($2.5$~\angstrom).\cite{art:3} (ii) In the right part of Fig.~\ref{fig: fig2}(c), a periodic corrugation is visible with a periodicity of $1.86 \pm 0.08$~nm and a corrugation of $0.45 \pm 0.05$~\angstrom. Considering that the moir\'{e} periodicity is 1.80 nm and the EMLG corrugation is 0.4~\angstrom{},\cite{art:6} we can confirm that in Fig.~\ref{fig: fig2}(c) the surface on the right is EMLG. (iii) Analyzing the left part of the image, close to the step, we obtain a corrugation in this area of $0.22 \pm 0.06$~\angstrom, which is only half of the value measured on EMLG.

In Fig.~\ref{fig: fig2}(b) we show an STM topographical image obtained after further Li deposition for a total of $0.047 \pm 0.003$~ML, which shows an inward extension of the flat areas with increasing Li amount. A similar behaviour was consistently observed in several areas on the sample. This is strong evidence that these areas are related to the Li deposition.

We can therefore exclude that these areas are bilayer inclusion. Three further reasons for this conclusion are: (i) We have not detected the presence of bilayer inclusions on this EMLG sample before Li deposition. (ii) The height difference between monolayer and bilayer graphene is $0.8 \pm 0.2$~\angstrom~\cite{art:3,art:8,Fiori2017,SH2017}, while here we measured an average height difference of $1.6 \pm 0.2$~\angstrom. (iii) Above bilayer graphene, the moir\'{e} pattern is visible, while here we observe a flat surface without any reconstruction.

The LEED data shown in Fig.~\ref{fig: fig1} provide clear evidence for Li--intercalation to the interface. Further evidence is provided by atomically resolved STM images. In the stripes close to the step edges, we observe the graphene structure with a measured periodicity of $2.43 \pm 0.05$~\angstrom. This data is in agreement with the well--known graphene lattice constant $a = 2.46$~\angstrom.\cite{art:1} This demonstrates that the topmost layer in this region is graphene. It also excludes the possible presence of Li clusters on the surface. This leaves the only possibility that Li is intercalated, transforming the EMLG into a QFBLG. This conclusion is further supported by the absence of the moir\'{e} pattern on the investigated surface. Furthermore, this process is observed to start at the SiC step edges, forming stripes which extend inward as the Li deposited amount increases. This is a strong evidence that Li--intercalation for the EMLG surface starts at the SiC step edges.

\begin{figure}[t]
\includegraphics[width=\columnwidth]{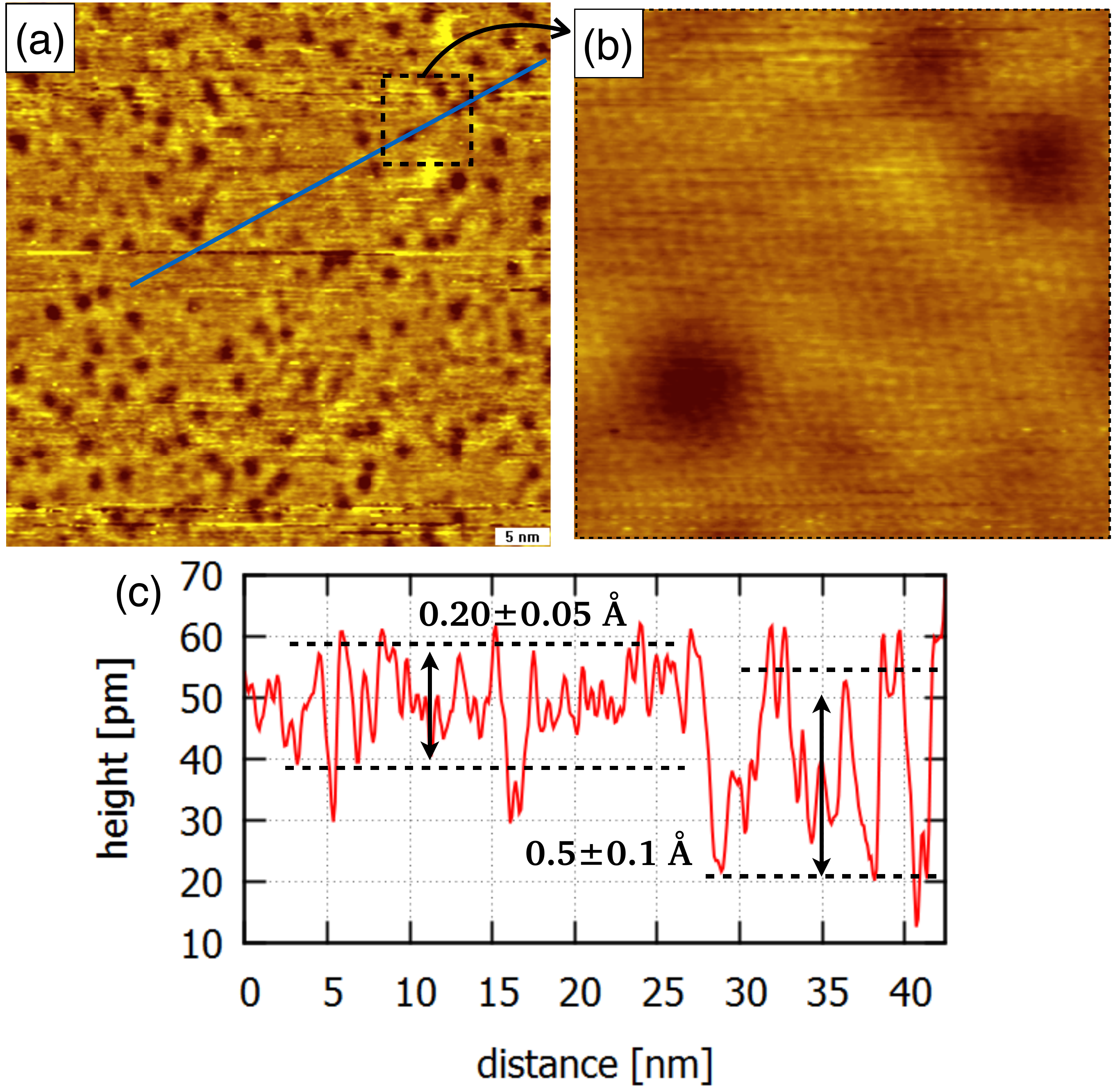}
\caption{\label{fig: fig3}EMLG surface after deposition of 0.28~ML Li. (a) STM topographical image. No moir\'{e} pattern is observed. Scan area: $50 \times 50$~nm$^2$. Image parameters: $-500$ mV, $-0.51$~nA. (b) Magnification ($6 \times 6$~nm$^2$) taken from the area indicated by the dashed square in (a). The graphene atomic lattice is resolved. Image parameters: 413 mV, 173 pA. (c) Cross sectional plot taken along the blue line in (a).}
\end{figure}

After deposition of 0.28~ML of Li on G/SiC, as we have shown in Fig.~\ref{fig: fig1}(b), the surface is $\left( 1 \times 1 \right)$ reconstructed, indicating the complete detachment of the buffer layer from the SiC substrate and its transformation to quasi--free--standing graphene. However, as shown in the STM image in Fig.~\ref{fig: fig3}(a), this surface does not appear perfectly flat, but it displays a random corrugation caused by dark spots without periodicity. Figure~\ref{fig: fig3}(b) shows a zoom--in STM image of Fig.~\ref{fig: fig3}(a) including some of these dark spots. The image clearly shows the graphene lattice also in the region of the spots, which allows us to confirm that the topmost surface layer is graphene, i.e.~these spots cannot be defects in the graphene. Furthermore, the cross sectional plot shown in Fig.~\ref{fig: fig3}(c), taken along the blue line in Fig.~\ref{fig: fig3}(a), shows that the depth of the dark spots is $0.5 \pm 0.1$~\angstrom, while the surface corrugation in the areas around the dark spots is $0.20 \pm 0.05$~\angstrom. This value is in good agreement with that reported in Fig.~\ref{fig: fig2}(d) for the intercalated area.

These data indicate an incomplete Li--intercalation at the interface. In detail, the parts of the surface in which the corrugation is equal to 0.20~\angstrom{} are intercalated areas in which the EMLG has been transformed to quasi--free--standing bilayer graphene (QFBLG). The dark spots could correspond to positions of Si atoms at the interface which are not saturated by Li. This causes the increase in surface corrugation to 0.5~$\angstrom$. It is likely that the Si atoms have dangling bonds, instead of covalent bonds with C atoms of the BL, because the depth of the dark spots of 0.5~$\angstrom$ is smaller than the step height (1.6~$\angstrom$) between QFBLG region intercalated with Li and EMLG region without Li, seen in Fig.~\ref{fig: fig2}. The observed surface structure is not new in literature for intercalated samples. Also for H--,\cite{Murata2014} Na--,\cite{art:109} and F--intercalated epitaxial graphene on SiC\cite{wong2015graphene} these surface features have been observed.

\begin{figure}[t]
\includegraphics[width=\columnwidth]{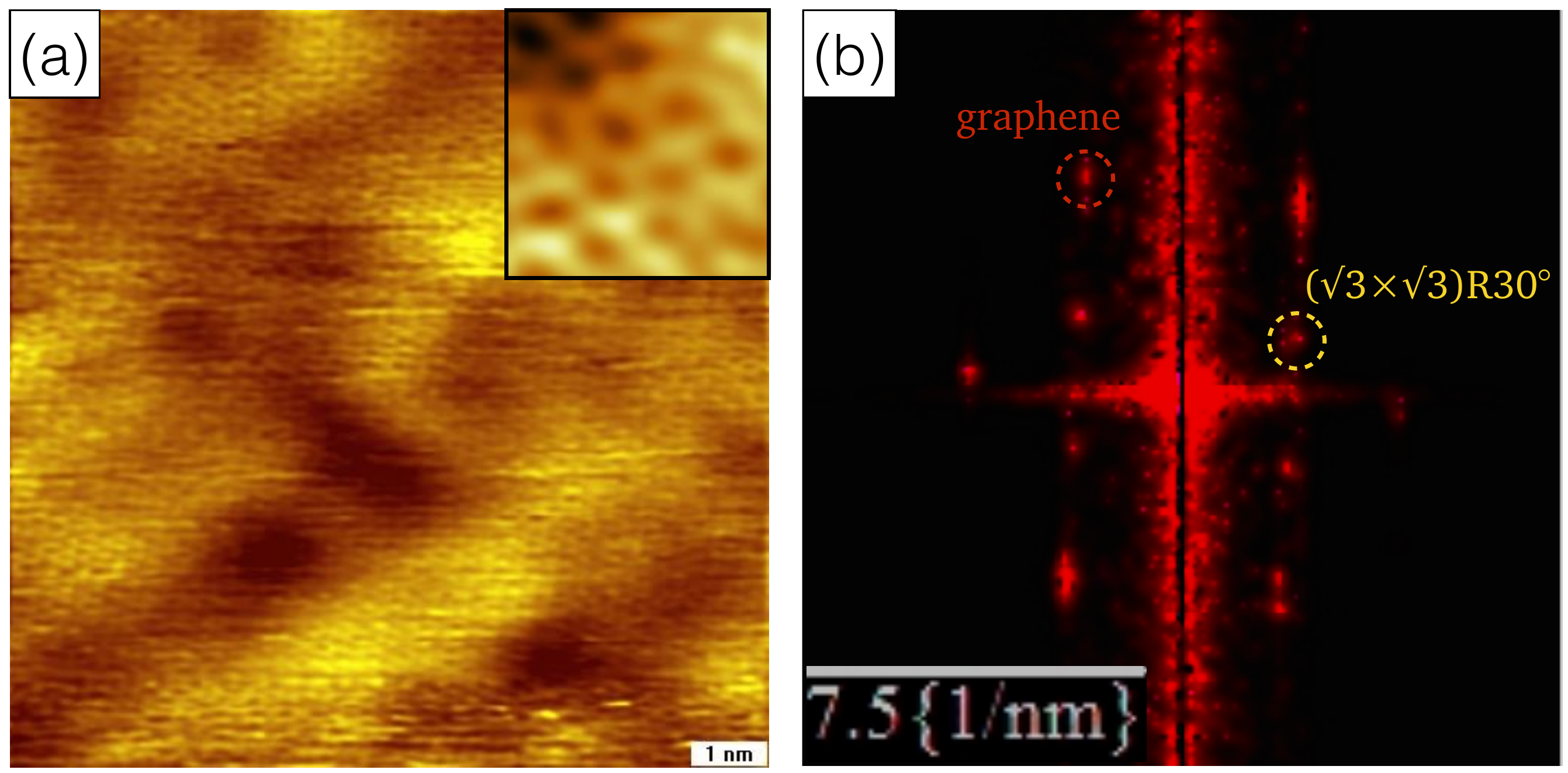}
\caption{\label{fig: fig4}EMLG surface after deposition of 0.56~ML Li. (a) STM image of this surface. Scan area: $10 \times 10$~nm$^2$.  The inset ($1 \times 1$~nm$^2$) shows a high--resolution image from the same area with atomic resolution. Image parameters: 30~mV, 90 pA. (b) 2D--FFT image of (a) in which the graphene spots and those of the $\left( \sqrt3 \times \sqrt3 \right)$R30$^\circ$ superstructure are well visible, highlighted by red and yellow circles, respectively. The scale bar indicates $k$ = 1 / wavelength.}
\end{figure}

From literature it is well known that about 30\% of the C--atoms of the buffer layer form covalent bonds to Si atoms of the SiC substrate, which corresponds to 0.3~ML.\cite{lin2015hydrogenation,sclauzero2012carbon} There is an excellent quantitative agreement between this number and the amount of Li that we had to deposit in order to induce the $\left( 1 \times 1 \right)$--reconstruction (0.28~ML), i.e.~for each Si atom at the interface, we deposited approximately one Li--atom. This suggests that complete intercalation will be obtained for a Li--coverage of 0.3~ML. Furthermore, this result allows for a quantitative comparison with existing intercalation models.\cite{art:78,art:88}

Finally, a further Li deposition for a total of 0.56~ML produces a $\left( \sqrt3 \times \sqrt3 \right)$R30$^\circ$ superstructure, as shown in Fig.~\ref{fig: fig1}(c). This superstructure has been associated in literature to intercalation between two graphene layers.\cite{art:16,art:15} In Fig.~\ref{fig: fig4}(a) we show an STM topographic image from a $10 \times 10$~nm$^2$ scan area from this surface. It appears clean and without Li clusters. In the inset, we show a high--resolution zoom--in of this surface with atomic resolution. The presence of the $\left( \sqrt3 \times \sqrt3 \right)$R30$^\circ$ superstucture in these images is confirmed by the FFT data of Fig.~\ref{fig: fig4}(a) which is shown in Fig.~\ref{fig: fig4}(b). In the latter, one $\sqrt3$ spot is indicated by the yellow circle.

\subsection{Li on the buffer layer}

\begin{figure}[t]
\includegraphics[width=\columnwidth]{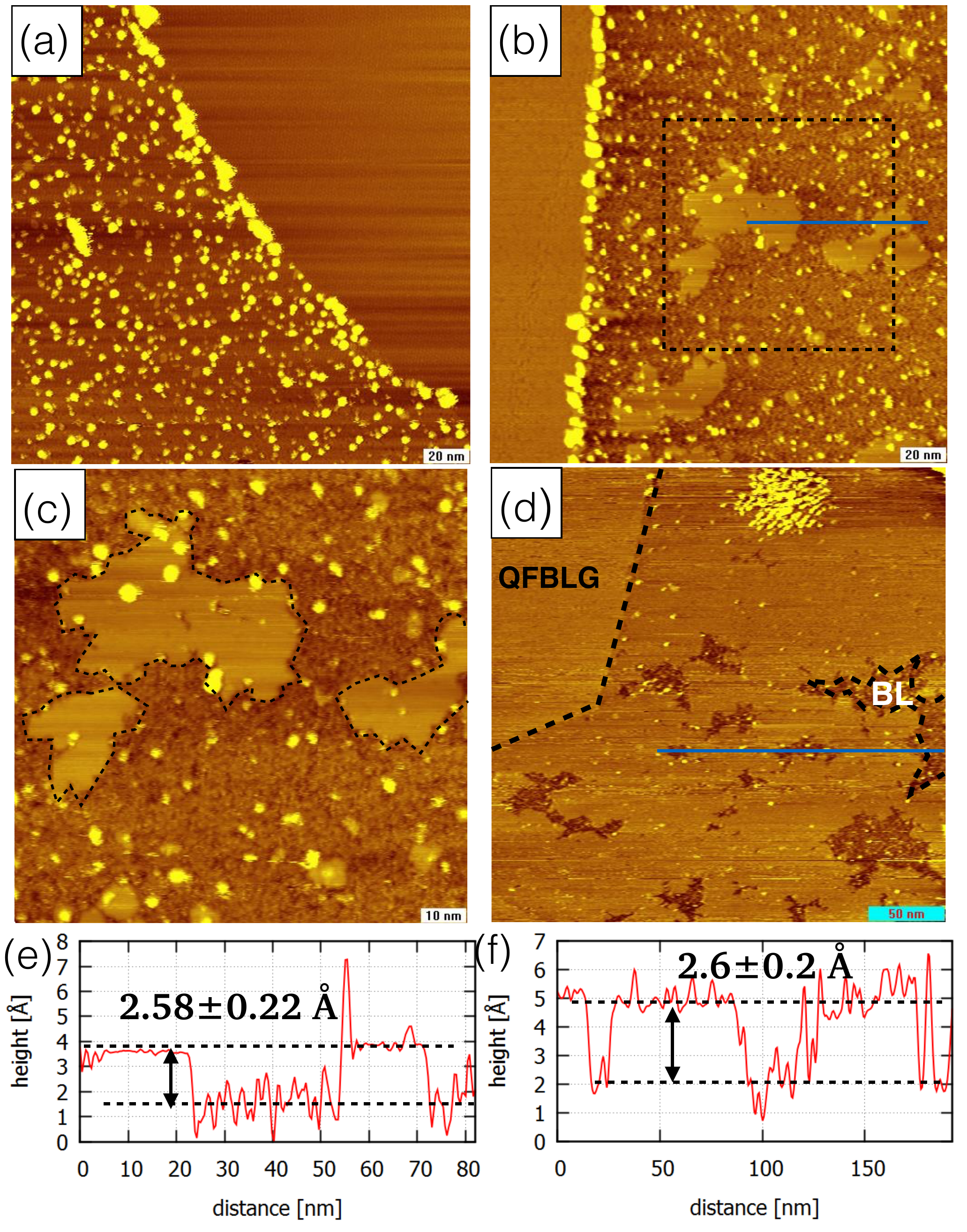}
\caption{\label{fig: fig5}(a) STM image from a BL area before Li deposition. The top-right shows an EMLG area. Scan area: $200 \times 200$~nm$^2$. (b) STM image from a BL area after deposition of 0.047~ML Li. On the left, a small (non-intercalated) EMLG area is visible. Scan area: $200 \times 200$~nm$^2$. (c) Magnification of the area indicated by the dashed square in (b). Scan area: $100 \times 100$~nm$^2$. Island formation in the BL area is visible. (d) STM topographic image from the BL after deposition of 0.28~ML Li. Scan area: $300 \times 300$~nm$^2$. The top--left shows a small QFBLG region. Most of the BL area has been transformed to QFMLG by the Li--intercalation. (e) Cross sectional plot taken along the blue line in (b), from which we can determine the islands' height. (f) Cross sectional plot taken along the blue line in (d). Images parameters: (a) 1 V, 1 nA; (b) and (c) $-300$ mV, $-1$~nA; (d) 0.8 V, 1~nA.}
\end{figure}

The presence of wide BL areas on our G/SiC samples allows to investigate the Li interaction with this carbon surface. It has already been experimentally demonstrated that BL and EMLG have the same lattice structure.\cite{art:7} However, due to the Si-C covalent bonds, the BL is more corrugated, and for this reason more reactive with other species. 

Fig.~\ref{fig: fig5}(a) shows a BL area before Li deposition, with an EMLG area in the top-right part. BL appears as decorated by yellow spots. These are carbon clusters formed during the growth process, due to the high reactivity of the BL surface. Indeed, the same features are not visible on the top-right EMLG area, which is smoother than the BL surface. These C clusters are still visible after Li deposition.

Our experimental STM observations on the BL demonstrate that Li evaporation (0.047~ML) at RT leads to the formation of islands. Figures~\ref{fig: fig5}(b)-(c) show an example. From the cross sectional profile reported in Fig.~\ref{fig: fig5}(e), the height of all islands is visibly the same ($2.6$~\angstrom) above a surface which shows a corrugation of approximately $1$~\angstrom, the value expected for the BL surface,\cite{art:7} while the surface of the islands is more flat. On top of them we can resolve the graphene lattice structure, but no moir\'{e} pattern, confirming that also in this case the Li is placed below the graphene. Our experimental STM observations therefore demonstrate that Li intercalates below the BL at RT, like for the EMLG surface. The islands have thus the same nature as the intercalated stripes along the step edges observed on the EMLG areas, which were discussed in Fig.~\ref{fig: fig2}.

Continuing the deposition of Li at RT up to 0.28~ML, the BL surface appears almost completely intercalated, as reported in Fig.~\ref{fig: fig5}(d). Therefore, most of the BL area has been transformed to QFMLG by the Li--intercalation. The top--left area of Fig.~\ref{fig: fig5}(d) shows a small QFBLG region, as identified by the absence of the moir\'{e} pattern, which indicates that also the EMLG has been intercalated and transformed. The cross--sectional profile displayed in Fig.~\ref{fig: fig5}(f), taken along the blue line in Fig.~\ref{fig: fig5}(d), shows that the height of the intercalated areas remains constant with increasing Li amount.

Although we observe Li intercalation both for EMLG and the BL, our experimental data highlight that the Li diffusion pathway to the interface in the BL case must be different from the EMLG case, where it occured via Li diffusion from the step edges. This will be discussed in more detail in section~\ref{discussion}.

\subsection{Annealing}

Annealing experiments performed on EMLG after deposition of 0.56 ML of Li at RT, provide important information on the stability of Li on this sample. Annealing to temperatures below 180 $^\circ$C does not change the surface order from a $\left( \sqrt3 \times \sqrt3 \right)$R30$^\circ$ superstructure, indicating that the compound is stable up to these temperatures. At this \lq\lq threshold temperature" (180 $^\circ$C), the LEED diffraction pattern changes to a $\left( 1 \times 1 \right)$ periodicity. This indicates that the Li which was placed between the BL and first layer graphene, is removed from this position, while the intercalated Li between the SiC substrate and the BL remains there. Therefore, the first important information which we collect is that there is a threshold in the use of Li--intercalated graphene compounds. This threshold temperature is around 200 $^\circ$C, and for higher temperatures the compound is irreparably modified. 

Increasing the annealing temperature, we collected LEED and STM evidence for a progressive Li desorption process from the G/SiC sample. The LEED results during annealing display the reverse process of that presented in Fig.~\ref{fig: fig1}: starting from a completely intercalated sample with a $\left( \sqrt3 \times \sqrt3 \right)$R30$^\circ$ superstructure, after several annealing cycles to always higher temperatures, the $\left( 1 \times 1 \right)$ and $\left( 6 \sqrt{3} \times 6 \sqrt{3} \right)$R30$^\circ$ reconstructions were progressively restored. The restoration of the moir\'{e} is clearly visible also by STM. Islands similar to those displayed in Fig.~\ref{fig: fig5}(c) appear in a kind of inverted intercalation process, and decrease in size and number as the annealing temperature increases. Finally, at about 600~$^\circ$C the Li desorption process appears completed, and this gives us a second important information: the Li deposition--intercalation process on G/SiC is reversible.

\section{\label{discussion}Discussion}

Let us now discuss about the nature of the terraces and islands on the EMLG and BL surfaces, respectively, analyzing more quantitatively the obtained STM results for both surface areas separately. By STM topographical images, it was possible to quantify how much the distance between the substrate and the BL is increased for the EMLG and BL surfaces due to the Li intercalation. 

\begin{figure}[t]
\includegraphics[width=\columnwidth]{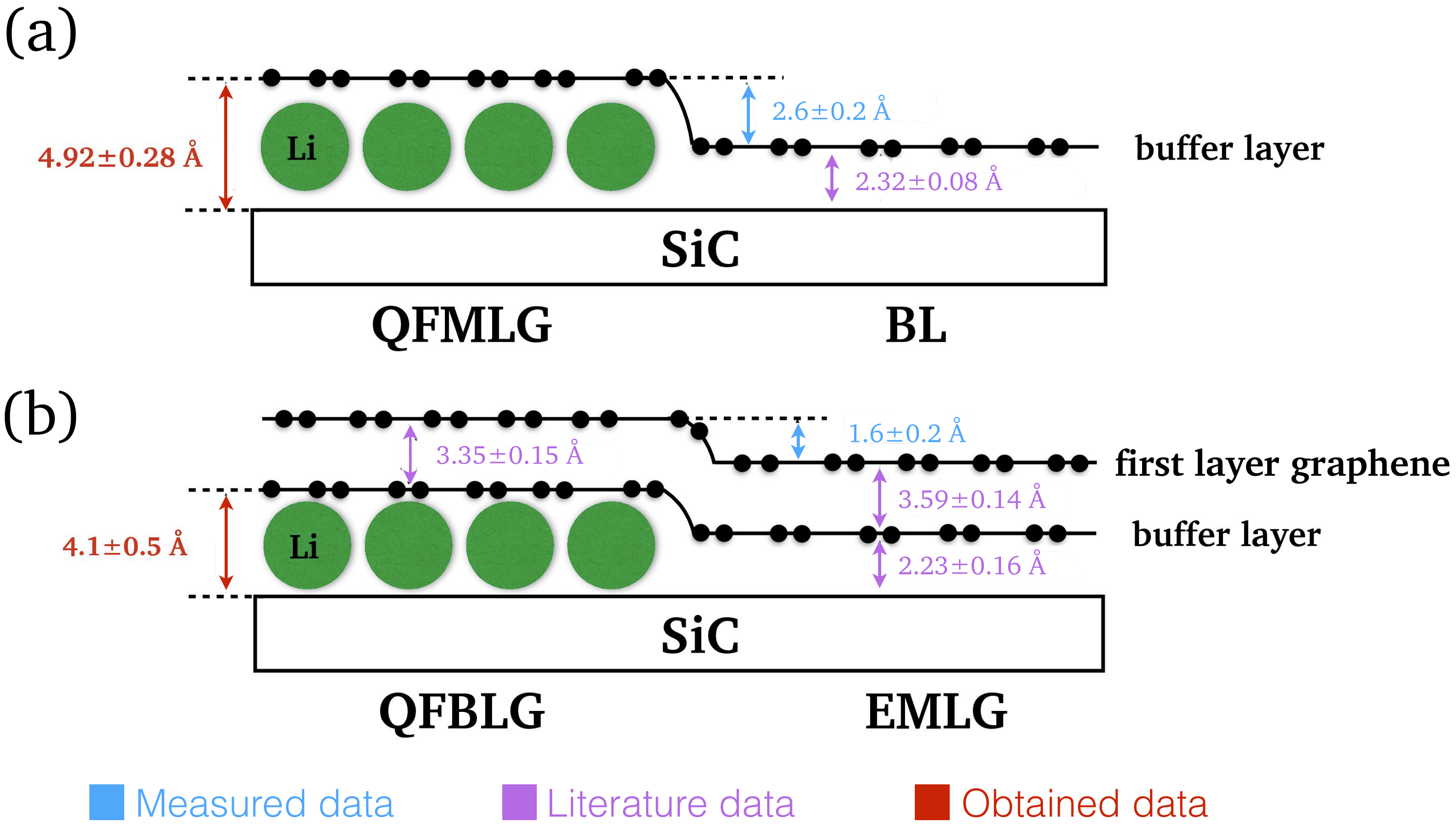}
\caption{\label{fig: fig6}Schematic representation of the Li distribution at the interface (a) for the BL case and (b) for the EMLG case. (a) Li converts the BL to QFMLG. The measured value of the island height in Fig.~\ref{fig: fig5}, $2.6$~\angstrom, allows to calculate the interlayer distance induced by the intercalated Li atoms between the substrate and the QFMLG. (b) Li converts EMLG to QFBLG. The measured value of the step height in Fig.~\ref{fig: fig2}, $1.6$~\angstrom, allows to calculate the interlayer distance induced by the intercalated Li atoms between the substrate and the QFBLG.}
\end{figure}

Regarding the BL areas, from literature, we know the distances between the substrate and the buffer layer ($2.32 \pm 0.08$~\angstrom).\cite{art:102} Through the STM topographical images, the islands height can be measured ($2.6 \pm 0.2$~\angstrom), as reported in Fig.~\ref{fig: fig5}. As shown in the sketch in Fig.~\ref{fig: fig6}(a), from this analysis we can deduce the interface distance between the substrate and the QFMLG as $4.92 \pm 0.28$~\angstrom. This result is in good agreement with theoretical predictions.\cite{art:78}

For the EMLG case, from literature, we know the distances between the substrate and the buffer layer ($2.23 \pm 0.16$~\angstrom)\cite{art:102} and between the buffer layer and the first graphene layer ($3.59 \pm 0.14$~\angstrom).\cite{art:102} Furthermore, in order to give an estimate of how much Li separates the region at the interface between substrate and BL, we need also to know the distance between a quasi--free--standing graphene and the first graphene layer above it. For this number, we considered useful the interlayer distance obtained for the H--intercalated system, since there are not yet numbers available for the Li--intercalated system. H--intercalation at the interface transforms the BL in a quasi--free--standing graphene which is spaced from the graphene layer above it by $3.35 \pm 0.15$~\angstrom.\cite{art:95} This number seems reasonable in comparison with the distances between various graphene layers. Using these numbers combined with our experimental results, we are able to calculate how much the BL is lifted up by the Li--intercalation at the interface. As reported in the sketch in Fig.~\ref{fig: fig6}(b), summing the measured data ($1.6 \pm 0.2$~\angstrom) with those available in literature for the distance between SiC substrate and BL, and BL and first layer graphene ($2.23 \pm 0.16$~\angstrom{} and $3.59 \pm 0.14$~\angstrom), then subtracting the distance between the detached BL and first layer graphene ($3.35 \pm 0.15$~\angstrom), we obtain a spacing of $4.1 \pm 0.5$~\angstrom{} between SiC substrate and the detached BL.

It is well-known that also the smallest atom of the periodic table, hydrogen, is not able to penetrate the graphene hexagons.\cite{Bunch2008,art:88} Thus, it is important to understand how Li can intercalate. Furthermore, it is important to keep in mind that typically intercalation is aided by heating, high pressure, etc. In this case, all the intercalation process occurs at RT.

For EMLG, we have shown that the Li--intercalation occurs mainly from the side of the SiC steps directly at the interface, without passing through the graphene mesh. Our experimental STM observations show intercalated stripes along the SiC step edges which extend more and more inward with increasing Li amount, as shown in Fig.~\ref{fig: fig2}. This clearly indicates that Li arrives at the interface mainly passing through the SiC step sides. This experimental result is in good agreement with other step intercalation mechanisms, already known in literature. \cite{petrovic2013mechanism,sicot2014copper,vlaic2014cobalt}

On the other hand, comparing the shape of the intercalated areas, one could think that Li intercalates differently in the BL case for which flat islands are observed inside the BL regions (see Fig.~\ref{fig: fig5}(b)). However, it should be considered that for the graphene growth conditions employed here (adopting a growth approach at atmospheric pressure) graphene grows from the SiC step edges inward, resulting in a larger BL presence in the middle of the terraces and not along the step edges. \cite{emtsev2009towards} For this reason, intercalation from step edges is unlikely in the BL case, making other paths (such as defects) more likely. Therefore, the presence of some defects in the BL can enhance the probability of Li to intercalate at the interface. In the monolayer case instead, the Li--intercalation through defects is more difficult because Li has to find two defects (one in the first layer graphene and one in the BL) before reaching the interface. For this reason, in the monolayer case, Li probably prefers to intercalate to the interface passing through the SiC step edges.

Our STM data indicate that for this intercalation process to occur, all sites at the SiC step edges are equally preferred, because we observe the formation of a rather homogeneous intercalated phase in stripe form along the SiC step edges. Initially, once a Li atom has reached the interface, it will with a high probability run into either a Si--C covalent bond or a Si dangling bond, and is captured to saturate the Si atom. This explains the formation of an intercalated stripe next to the SiC step edge. The formation of this stripe might also promote further intercalation, since, as we have shown, it leads to an increase in layer spacing between the SiC substrate and the BL. For higher Li coverages, since all available Si atoms close to the setp edge are already saturated, the Li atom has to diffuse further inward, but also in this case it is captured as soon as it hits a Si atom not yet saturated with Li. In this way the intercalated stripe progressively grows inwards until the whole terrace is intercalated.

If we compare these results to the well--studied case of H--intercalation of G/SiC, we note some differences. Hydrogen intercalation performed under typical process conditions (H--pressure 1013~mbar and temperature $600 - 1200~^\circ$C)\cite{Murata2014} occurs almost instantaneous and complete.\cite{Riedl2010} This means that once a path is created (either at the step edge or at a defect site), the entire terrace intercalates at once. However, we have to consider two factors. Hydrogen intercalation occurs from the gas phase, with an approximately infinite supply of hydrogen. Here, instead, we supply a limited (and well--known) submonolayer--amount of Li at a time, so there is simply not enough Li available initially to intercalate the whole terrace at once. We note that islanding has also been reported for H--intercalation under non--standard UHV conditions.\cite{Watcha2011} Second, our experiments are performed at room temperature, while H--intercalation requires elevated temperatures ($600 - 1200~^\circ$C). It can be expected that temperature affects kinetics, and so a limited diffusivity of Li might be one of the reasons for the observation of intercalated stripes and islands for EMLG and BL, respectively, in our case. We stress that intercalation temperature has also a strong influence on the kinetics of H--intercalation.\cite{Deretzis2013,Murata2014} Therefore, Li--intercalation experiments at higher temperatures might be able to shine further light on these issues.

\section{Conclusions}

We have investigated the Li deposition on epitaxial graphene on silicon carbide by STM and LEED techniques in UHV conditions. STM topographical images show the progressive Li--intercalation on EMLG and BL surfaces. Through a diffusion process, Li is shown to move to the interface, where it breaks the Si--C covalent bonds and saturates the Si dangling bonds between substrate and BL. The latter is in this way lifted up and transformed into quasi--free--standing graphene. For the case of EMLG, the STM data indicates that Li arrives at the interface directly from the SiC step edges. A quantitative analysis made on EMLG and BL surfaces allowed to measure the distance between the SiC substrate and the BL induced by Li--intercalation. Once the intercalation process at the interface is completed, Li--intercalation continues in between the two graphene layers. This is demonstrated by the observation of the $\left( \sqrt3 \times \sqrt3 \right)$R30$^\circ$ superstructure on the EMLG surface by LEED and STM. Finally, several annealing cycles performed on Li--intercalated graphene reverse the intercalation process and induce the Li desorption from the G/SiC sample.
 
\begin{acknowledgments}
Funding from the European Union Seventh Framework Programme under Grant Agreement No. 696656 Graphene Core 1 is acknowledged. Financial support from the CNR in the framework of the agreements on scientific collaborations between CNR and CNRS (France), NRF (Korea), and RFBR (Russia) is also acknowledged. We further acknowledge funding from the Italian Ministry of Foreign Affairs, Direzione Generale per la Promozione del Sistema Paese, and from the Polish Ministry of Science and Higher Education, Department of International Cooperation (agreement on scientific collaboration between Italy and Poland). We also acknowledge support from Scuola Normale Superiore, project SNS16\_B\_HEUN – 004155.
\end{acknowledgments}

\bibliography{biblio-sh}

\end{document}